# Deconvolution of a linear combination of Gaussian kernels by Liouville-Neumann series applied to an inhomogeneous Fredholm integral equation of second kind and applications to image processing

W. Ulmer

Klinikum München-Pasing, Department of Radiation Therapy and MPI of Physics, Göttingen, Germany

e-mail: waldemar.ulmer@gmx.net

**Abstract**

Scatter processes of photons lead to blurring of images produced by CT (computed tomography) or CBCT (cone beam computed tomography) in the KV domain or portal imaging in the MV domain (KV: kilovoltage, MV: megavoltage). Multiple scatter is described by, at least, one Gaussian kernel. In various situations, this approximation is crude, and we need two/three Gaussian kernels to account for the long-range tails, appearing in the Molière scatter of protons or in Compton scatter of photons.  If image structures are obtained by measurements, these structures are always blurred by scattering. The ideal image (source function) is subjected to Gaussian convolutions to yield a blurred image recorded by a detector array. The inverse problem is to obtain the ideal source image from measured image. A new deconvolution method for linear combinations of two/three Gaussian kernels with different parameters $s_0$, $s_1$, $s_2$ is derived via an inhomogeneous Fredholm integral equation of second kind (IFIE2) and Liouville - Neumann series (LNS) to provide the source function ρ. A comparison with previously published results is the main purpose in this study. We can verify advantages of this method in image processing applied to inverse problems (two/three kernels) of CBCT or IMRT (intensity-modulated radiotherapy) detector arrays of portal imaging. A particular advantage of this procedure is given, if the scatter functions $s_0$, $s_1$, $s_2$ are not constant and depend on coordinates. This fact implies that the scatter functions can be calibrated according to the electron density $\rho_{electron}$ provided by image reconstructions.

## 1. Introduction

Various scatter processes of photons lead to blurring of images produced by CT/CBCT or portal imaging (KV/MV domain). Multiple scatter can be described by, at least, one single Gaussian kernel [1 - 3], which we formally abbreviate by K(s, u – x), but it may refer to more than one dimension. The ideal image (source function ρ without any blurring) is subjected to a Gaussian convolution in order to yield a blurred image φ, which may be recorded by a detector array:

$$K = \frac{1}{s \cdot \sqrt{\pi}} \cdot \exp(-(u-x)^2/s^2)$$
$$\varphi = \int K(s, u-x) \cdot \rho(u) du \qquad (1)$$

In equation (1), we may refer to $\rho$ as a source function and to $\varphi$ as an image function. The magnitude of the parameter s represents a measure of the severeness of the blurring, since by taking $s \rightarrow 0$ the kernel K assumes the shape of a $\delta$-function and $\varphi$ becomes identical with $\rho$.

In many situations [5 - 6] the restriction to one Gaussian kernel represents a crude approximation, and we need a linear combination of Gaussian kernels with $K_g$ as a resulting convolution kernel to account for long-range tails, which appear in the Molière multiple scatter theory of protons/electrons or in Compton scatter of $\gamma$-quanta:

$$K_g = c_0 \cdot K_0 + c_1 \cdot K_1 + c_2 \cdot K_2$$
$$K_j = \frac{1}{s_j \cdot \sqrt{\pi}} \cdot \exp(-(u-x)^2/s_j^2)$$
$$j = 0, 1, 2 \qquad c_0 + c_1 + c_2 = 1$$
$$\varphi = \int K_g(s_0, s_1, s_2, c_0, c_1, c_2, u-x) \cdot \rho(u) du \qquad (2)$$

The parameters in equation (2) have to satisfy $c_0 > (c_1, c_2)$ and $s_0 < s_1 < s_2$. The restriction to two Gaussian kernels results by setting $c_2 = 0$.

The inverse problem of this procedure is to calculate the ideal source image from really determined image. If the scatter parameters are known (e.g. *rms* value s of Gaussian kernels via appropriate test measurements or Monte-Carlo simulations), we are able to calculate the idealistic source structure by an inverse kernel $K^{-1}(s, u - x)$:

$$\rho(x) = \int K^{-1}(s, u-x) \cdot \varphi(u) du. \qquad (3)$$

In the meantime, the inverse kernel $K^{-1}(s, u-x)$ of a single Gaussian kernel $K(s, u - x)$ is a well-established, rigorous mathematical tool avoiding ill-posed aspects. There are two possible representations of the inverse kernel $K^{-1}(s, u - x)$:

$$K^{-1}(s, u-x) = \sum_{n=0}^{N} c_n(s) \cdot H_{2n}\left(\frac{u-x}{s}\right) \cdot K(s, u-x)$$
$$N \rightarrow \infty; \quad c_n = (-1)^n \cdot s^{2n}/(2^n \cdot n!); \quad (n = 0, 1,..., \infty) \qquad (4)$$

$$K^{-1}(s, u-x) = \delta(u-x) + \sum_{n=1}^{N} c_n(s) \cdot H_{2n}\left(\frac{u-x}{s}\right) \cdot K(s, u-x)$$
$$N \to \infty; \quad c_n(s) = (-1)^n \cdot s^{2n} \cdot (2^n - 1)/(4^n \cdot n!); \quad (n = 1, \ldots, \infty) \qquad (5)$$

In both representations, $H_{2n}$ refer to Hermite polynomials of even order. Therefore the inverse kernel $K^{-1}$ can be regarded as a generalized Gaussian convolution kernel, since the kernel $K^{-1}$ contains two-point Hermite polynomials $H_{2n}((u - x)/s)$ in a rigorously defined order. In practical applications, it is necessary to restrict N to a finite limit ($N < \infty$), and the question arises, which N provides sufficiently accurate results. Based on formulas (4 – 5) there have been put forward many applications in radiation physics, mainly with regard to the deconvolution problem of the finite detector size in radiation physics [6 – 10]. It should also be noted that the simplest, but well-known solution function of the heat/diffusion equation is a Gaussian kernel [11 – 12]. The inverse problem of this distribution function [13 – 15] is similar to the problem given by equation (3); it represents a typical case of an ill-posed problem and requires regularizations techniques, which have been applied by the aforementioned authors. However, it appears that in this field the EM algorithm [16 – 18] has proven to be valuable.

We shall now extend our considerations to the inverse problem of a linear combination of two/three Gaussian convolution kernels $K_g^{-1}(s_0, c_0, s_1, c_1, s_2, c_2, u - x)$ according to equation (2), in order to found applications to aforementioned image processing, where a single Gaussian kernel would represent a crude approximation. The kernels $K_g(s_0, c_0, s_1, c_1, s_2, c_2, u - x)$ and $K_g^{-1}(s_0, c_0, s_1, c_1, s_2, c_2, u - x)$ account for those situations, where small long-range tails are present, and the restriction to one kernel K and its inverse kernel $K^{-1}$ is apparently deficient. In this communication, we shall develop a new solution procedure of the inverse problem of a linear combination of two Gaussian kernels which avoids the determination of the deconvolution kernel $K_g^{-1}$, namely its formulation by an IFIE2 and related LNS to calculate solutions in every desired order. The results obtained by the LNS procedure will be compared with a different procedure to calculate $K_g^{-1}$ from $K_g$, which has been previously published. With regard to applications we preferably consider problems of image processing in the KV and MV domain. In the next section, the inverse kernel $K_g^{-1}$ will be presented and developed according to an IFIE2 and LNS procedure; it represents a tool in IMRT and intensity-modulated proton therapy (IMPT), see e.g. [19 - 21]. We should also point out that in many problems of deconvolutions fast Fourier transforms (FFT) together with Wiener filters are applied. A very concise paper on Fourier-based deconvolutions and filter functions has been given in a review paper [22]. However, some critical aspects result from Fourier-based deconvolutions applied to step functions and are usually referred to as 'ill-posed' (see the applications given in a later section). These well-known problems have previously been discussed [6, 23 – 24]. Since Gaussian-like convolutions/deconvolutions play a significant role in many disciplines of physics and engineering [1 – 10, 19 – 27], reliable toolkits for inverse procedures are desirable, which circumvent ill-posed aspects.

## 2. Methods

### 2.1. Operator calculus (Lie series of operators) and the derivation of the inverse kernels

At first, we shortly summarize previous results [2, 6], which should be consulted by those readers with need of more detailed information. A very convenient way is the operator formulation to derive the Gaussian convolution kernel as a Green's function and the related inverse problem.

The basic formulas of all subsequent procedures and calculations are the following two operator functions:

$$O = exp(-\tfrac{1}{4} \cdot s^2 \cdot \tfrac{d^2}{dx^2}) \quad and \quad O^{-1} = \exp(\tfrac{1}{4} \cdot s^2 \cdot \tfrac{d^2}{dx^2}). \qquad (6)$$

In three dimensions, we have to substitute operator $d^2/dx^2$ by the 3D Laplace operator $\Delta$:

$$O = \exp(-\tfrac{1}{4} \cdot s^2 \cdot \Delta) \quad and \quad O^{-1} = \exp(\tfrac{1}{4} \cdot s^2 \cdot \Delta). \qquad (7)$$

O and $O^{-1}$ obey the following relation:

$$\left.\begin{array}{l} O \cdot O^{-1} = O^{-1} \cdot O = \exp(-\tfrac{1}{4} \cdot s^2 \cdot \tfrac{d^2}{dx^2}) \cdot \exp(\tfrac{1}{4} \cdot s^2 \cdot \tfrac{d^2}{dx^2}) \\ = \exp\left(-\tfrac{1}{4} \cdot s^2 \cdot \tfrac{d^2}{dx^2} + \tfrac{1}{4} \cdot s^2 \cdot \tfrac{d^2}{dx^2}\right) = 1 \; (unit\; operator) \end{array}\right\}. \qquad (8)$$

The operators O and $O^{-1}$ and their related actions to a class of functions are formally defined by Taylor series of the exponential functions, which represent Lie series of operator functions [2]:

$$O^{-1} = 1 + \sum_{n=1}^{\infty} \frac{s^{2n}}{n! \cdot 4^n} \cdot d^{2n}/dx^{2n}; \quad O = 1 + \sum_{n=1}^{\infty} \frac{s^{2n}}{n! \cdot 4^n} \cdot (-1)^n \cdot d^{2n}/dx^{2n}. \qquad (9)$$

If $\rho(x)$ represents a source and $\varphi(x)$ an image function, the following relationships are obtained:

$$\varphi(x) = O^{-1} \cdot \rho(x); \; \rho(x) = O \cdot \varphi(x). \qquad (10)$$

It can be concluded from relations (6 – 10) that all permitted functions ρ(x) and φ(x) have to belong to the function space $C^\infty$ (Banach space), which implies that both sets φ(x) and ρ(x) are defined by derivatives of infinite order. The integral operator notation (Green's function) of $O^{-1}$ and O are [2]:

$$\left. \begin{array}{l} \varphi(x) = O^{-1} \cdot \rho(x) = \int K(s, u-x) \cdot \rho(u) du \\ \rho(x) = O \cdot \varphi(x) = \int K^{-1}(s, u-x) \cdot \varphi(u) du \end{array} \right\}. \qquad (11)$$

The integral operator kernel K is the normalized Gaussian kernel according to equation (1), which is based on the spectral theorem of functional analysis [2]. The essential difference between the differential and integral operator formulation is the class of the permitted functions. It should be noted that the equivalence of the two different calculation methods is only ensured for special class of functions we denote by $C^\infty$. Thus in the case of the differential operator functions O and $O^{-1}$ acting on φ and ρ the restriction to $C^\infty$ is a necessary and sufficient condition (finite-order polynomials $P_n(x) = a_0 + a_1 x + a_2 x^2 + \ldots + a_n x^n$ are included), and the action of the operator O does not lead to an ill-posed operation and to a necessary regularization. An ill-posed problem appears, which we have to circumvent, if the operator O acts on Fourier transforms, whereas with regard to the operator $O^{-1}$ this problem does not emerge. For this purpose we consider Fourier expansions of the functions φ and ρ, and the action of the operators O and $O^{-1}$ on them.

$$\rho = \frac{1}{\sqrt{2\pi}} \int \rho'(k) \cdot \exp(i \cdot k \cdot x) dk \quad \text{and} \quad \varphi = \frac{1}{\sqrt{2\pi}} \int \varphi'(k) \cdot \exp(i \cdot k \cdot x) dk. \quad (12)$$

The operation $O^{-1} \cdot \rho$ provides:

$$O^{-1} \cdot \rho = \frac{1}{\sqrt{2\pi}} \int \exp(0.25 \cdot s^2 \cdot d^2/dx^2) \rho'(k) \cdot \exp(i \cdot k \cdot x) dk = \frac{1}{\sqrt{2\pi}} \int \rho'(k) \cdot \exp(i \cdot k \cdot x) \cdot \exp(-0.25 \cdot s^2 \cdot k^2) dk. \quad (12a)$$

According to previous results [2, 11 – 12] the Green's function associated with the operator $O^{-1}$ can be derived from the right-hand side of the above equation (12a) with the help of the spectral theorem of functional analysis and turns out to be the normalized Gaussian kernel (1):

$$K(s, u-x) = \frac{1}{\sqrt{2\pi}} \frac{1}{\sqrt{2\pi}} \int \exp(-i \cdot k \cdot u) \cdot \exp(i \cdot k \cdot x) \cdot \exp(-0.25 \cdot s^2 \cdot k^2) dk. \quad (12b)$$

We now perform the identical procedure with respect to the operator O:

$$O^{-1} \cdot \varphi = \frac{1}{\sqrt{2\pi}} \int \exp(0.25 \cdot s^2 \cdot d^2/dx^2) \varphi'(k) \cdot \exp(i \cdot k \cdot x) dk = \frac{1}{\sqrt{2\pi}} \int \varphi'(k) \cdot \exp(i \cdot k \cdot x) \cdot \exp(0.25 \cdot s^2 \cdot k^2) dk. \quad (12c)$$

It is easy to verify that the right-hand side of equation (12c) does in general not exist due to the term $\exp(0.25 \cdot k^2 \cdot s^2)$, and it can only be regularized, if $\varphi'(k)$ vanishes sufficiently fast. In particular, the Green's function related to the operator O cannot be derived from the analogue expression of formula (12b). In order to obtain the integral operator kernel $K^{-1}$ of the operator O, which works without any restrictions, we write:

$$O = \exp(-0.25 \cdot s^2 \cdot d^2/dx^2) = \exp(-0.5 \cdot s^2 \cdot d^2/dx^2) \cdot \exp(0.25 \cdot s^2 \cdot d^2/dx^2). \quad (13)$$

Thus the operator $\exp(0.25 \cdot s^2 \cdot d^2/dx)$ on the right-hand side of the above equation (13) yields the kernel K(s, u - x). The action of the operator $\exp(-0.5 \cdot s^2 \cdot d^2/dx^2)$ is obtained by its Lie series acting now on K:

$$\exp(-0.5 \cdot s^2 \cdot d^2/dx^2) \cdot K(s, u-x) = \sum_{n=0}^{\infty} (-1)^n \cdot 2^{-n} \cdot \frac{1}{n!} \cdot d^{2n}/dx^{2n} \cdot K(s, u-x). \quad (13a)$$

Using the definition of Hermite polynomials, we obtain the relations (4, 5) from the right-hand side of equation (13a).

The integral operators K and $K^{-1}$ only require the Banach space $L_1$ of Lebesque-integrable functions. This fact has an important meaning in practical applications, where summations in finite intervals have to be accounted for (step functions, voxel integrations). The inverse integral operator kernel $K^{-1}$ has already been presented by equations (3 – 5). The integral operator correspondence to equation (8), i.e. $O \cdot O^{-1} = 1$, is given by the following equation:

$$\int K(s, u-u') \cdot K^{-1}(s, u'-x) du' = \delta(u-x). \quad (14)$$

It has to be mentioned that relation (14) is also valid for every kind of integral operators, if both K and $K^{-1}$ exist. There are various problems, where the differential operator calculus is easier to handle, e.g. the derivation of basic formulas, and we mention the properties of iterated kernels. The repeated application of the operators $O^{-1}$ and O (*n times*) leads to the expressions:

$$\left.\begin{array}{l} O^{-1} \cdot \ldots \cdot O^{-1} (n\ times) = O^{-n} = \exp\left(\frac{s_n^2}{4} \cdot \frac{d^2}{dx^2}\right) \\ O \cdot \ldots \cdot O\ (n\ times) = O^n = \exp\left(-\frac{s_n^2}{4} \cdot \frac{d^2}{dx^2}\right) \\ s_n^2 = n \cdot s^2 \end{array}\right\}. \quad (15)$$

The integral operator kernels of relation (15) are simply given by the modification:

$$\left. \begin{array}{l} K(s, u-x) \Rightarrow K(s_n, u-x); \quad K^{-1}(s, u-x) \Rightarrow K^{-1}(s_n, u-x) \\ s^2 \Rightarrow s_n^2 = n \cdot s^2 \end{array} \right\}. \quad (16)$$

The 3D extension of the relation (1) is the 3D Gaussian convolution kernel, which reads:

$$K(s, u-x, v-y, w-z) = \frac{1}{\sqrt{\pi}^3} \cdot \frac{1}{s^3} \cdot \exp\left(-\frac{1}{s^2} \cdot ((u-x)^2 + (v-y)^2 + (w-z)^2)\right). \quad (17)$$

Integrations have to be carried out over $u$, $v$ and $w$. The integral operator correspondence $K^{-1}$ of equation (17) is obtained in a similar way. For this purpose, we write the Hermite polynomial expansion in each dimension according to equation (4) by introducing the terms $F_1(s, u-x)$, $F_2(s, v-y)$ and $F_3(s, w-z)$. $F_1$, $F_2$ and $F_3$ will be defined as below and have to be multiplied with the kernel K according to equation (17). By that, we obtain:

$$\left. \begin{array}{l} F_1(s, u-x) = \sum_{n=0}^{N} c_n(s) \cdot H_{2n}(\frac{u-x}{s}); \quad F_2(s, v-y) = \sum_{n=0}^{N} c_n(s) \cdot H_{2n}(\frac{v-y}{s}); \quad F_3(s, w-z) = \sum_{n=0}^{N} c_n(s) \cdot H_{2n}(\frac{w-z}{s}) \\ N \to \infty \end{array} \right\}. \quad (18)$$

$$K^{-1}(s, u-x, v-y, w-z) = \prod_{k=1}^{3} F_k \cdot K(s, u-x, v-y, w-z). \quad (19)$$

Equation (19) is also valid, if the substitution $s^2 \to s_n^2 = n \cdot s^2$ is performed. In the preceding section we have stated arguments, why in some situations a linear combination of Gaussian convolution kernels is required according to equation (2). The operator notation analog to equation (6) of this convolution reads (in one dimension):

$$\varphi(x) = O_g^{-1} \cdot \rho(x) = [c_0 \cdot O_0^{-1}(s_0) + c_1 \cdot O_1^{-1}(s_1) + c_2 \cdot O_2^{-1}(s_2)] \cdot \rho(x). \quad (20)$$

It is the task to determine $\rho(x)$, if $\varphi(x)$ is given; this is equivalent to the determination of the kernel $K_g^{-1}$. In every case, the condition $O_g^{-1} \cdot O_g = 1$ has to be satisfied.

In a previous study [2] we have made use of the operator calculus to determine $O_g$ and via $O_g^{-1}$ to derive the inverse kernel $K_g^{-1}$. The operator calculus provides the following relationship:

$$\left. \begin{array}{l} O_g \cdot O_g^{-1} = 1 = O_g \cdot [c_0 \cdot O_0^{-1} + c_1 \cdot O_1^{-1} + c_2 \cdot O_2^{-1}] \\ \Rightarrow O_g = [c_0 \cdot O_0^{-1} + c_1 \cdot O_1^{-1} + c_2 \cdot O_2^{-1}]^{-1} \end{array} \right\}. \quad (22)$$

We have now to evaluate the following Lie series of the operator function $[c_0 \cdot O_0^{-1} + c_1 \cdot O_1^{-1} + c_2 \cdot O_2^{-1}]^{-1}$ in terms of the operators $O_0^{-1}$ and $O_1^{-1}$. We use the following relation for commutative operators:

$$[A+B]^{-1} = \sum_{n=0}^{\infty} (-1)^n \cdot A^{-n-1} \cdot B^n. \quad (23)$$

With the help of the substitutions $A = c_0 \cdot O_0^{-1}$ and $B = c_1 \cdot O_1^{-1} + c_2 \cdot O_2^{-1}$ we are able to derive the operator function $O_g$, which satisfies $O_g \cdot \varphi = \rho$, and the related inverse kernel $K_g^{-1}$:

$$\varphi = [c_0 \cdot O_0^{-1} + c_1 \cdot O_1^{-1} + c_2 \cdot O_2^{-1}] \cdot \rho. \quad (24)$$

$$\Rightarrow \frac{1}{c_0} \cdot O_0 \cdot [1 + \frac{c_1}{c_0} \cdot O_0 \cdot O_1^{-1} + \frac{c_2}{c_0} \cdot O_0 \cdot O_2^{-1}]^{-1} \cdot \varphi = \rho. \quad (25)$$

With the help of equation (23) the integral operator notation of equation (25) assumes the shape:

$$\left.\begin{array}{l} K_g^{-1} = \frac{1}{c_0} \cdot K_0^{-1} + \sum_{n=1}^{M}(-1)^n \cdot c_0^{-n-1} \cdot \sum_{j=0}^{n} \binom{n}{j} \cdot c_1^{n-j} \cdot c_2^j \cdot K(\tau_{n,j}, u-x) \\ \tau_{n,j}^2 = s_1^2 \cdot (n-j) + s_2^2 \cdot j - (n+1) \cdot s_0^2; \quad M \to \infty \end{array}\right\} \quad (26)$$

It can immediately be concluded from equations (24 - 26) that the following conditions must hold:

(1) The case n = j = 0 yields the operator function $\exp(-0.25 \cdot s_0^2 \cdot d^2/dx^2)/c_0$ in relation (25), i.e. we obtain in every case a deconvolution kernel of the type (4).

(2) The case n = 1, j = 0 provides the following term $\tau_{1,0}^2 = s_1^2 - 2 \cdot s_0^2$ in the exponential operator function. In order to obtain a convolution term induced by $\exp(0.25 \cdot \tau_{1,0}^2 \cdot d^2/dx^2)$ the condition $\tau_{1,0}^2 = s_1^2 - 2 \cdot s_0^2 > 0$ has to be satisfied. We assume that this condition is satisfied; then it is also satisfied for all terms with n > 1 yielding $\tau_{n,j}^2 = (n-j) \cdot s_1^2 + j \cdot s_2^2 - (n+1) \cdot s_0^2 > 0$. This implies that all terms of the expansion (26) refer to convolution expansions, which have to be evaluated. We refer this case as to the *standard case*, otherwise the expansion (26) has to be treated in a rather different way. The so-called Mexican hat with $s_0 > s_1$, $c_1 < 0$, $c_2 = 0$, $c_0 + c_1 = 1$ might be a noteworthy example [2].

For completeness, we note the integral operator $K_g^{-1}$ in the case of two Gaussian kernels ($c_0 + c_1 = 1$):

$$\left.\begin{array}{l} K_g^{-1}(c_0, s_0, c_1, s_1, u-x) = \frac{1}{c_0} \cdot K_0^{-1}(s_0, u-x) + \frac{1}{c_0} \cdot \sum_{n=1}^{M}(-1)^n \cdot (c_1/c_0)^n \cdot K(n \cdot s_1^2 - (n+1) \cdot s_0^2, u-x) \\ M \to \infty \end{array}\right\} \quad (27)$$

Equation (26) may readily be extended to 3D:

$$\left.\begin{array}{l} K_g^{-1}(c_0, s_0, c_1, s_1, c_2, s_2, u-x, v-y, w-z) = \frac{1}{c_0} \cdot K_0^{-1}(s_0, u-x, v-y, w-z) + \\ + \frac{1}{c_0} \cdot \sum_{n=1}^{M}(-1)^n \cdot c_0^{-n-1} \cdot \sum_{j=0}^{n} \binom{n}{j} \cdot c_1^{n-j} \cdot c_2^j \cdot K(\tau_{n,j}, u-x, v-y, w-z) \\ \tau_{n,j}^2 = (n-j) \cdot s_1^2 + j \cdot s_2^2 - (n+1) \cdot s_0^2; \quad N_f = \frac{1}{\sqrt{\pi}^3} \cdot \frac{1}{\tau_{n,j}^3}; \quad M \to \infty \end{array}\right\} \quad (28)$$

The term $K_0^{-1}$ in equations (26 – 28) has to be treated according equations (4 - 5) or equation (19) in the case of three dimensions. In view of the following section, we introduce the abbreviation:

$$f(x,y,z) = \frac{1}{c_0} \cdot \int K_0^{-1}(s_0, \vec{u} - \vec{x}) \cdot \varphi(\vec{u}) d^3u. \quad (29)$$

Function $f$ incorporates the inhomogeneous part of the Fredholm integral equation of second kind (IFIE2).

*2.2. Inverse problem according to IFIE2 and LNS method*

In order to derive an alternative method to solve the inverse problem of a linear combination of Gaussian convolutions, we consider equation (20) with regard to two kernels (the generalization to $c_2 \neq 0$ will be stated thereafter), which we multiply with $O_0/c_0$ from the left-hand side. By that, we readily obtain the desired formula, which will be transformed to a IFIE2:

$$\frac{1}{c_0} \cdot O_0 \cdot \varphi = [1 + \frac{c_1}{c_0} \cdot O_0 \cdot O_1^{-1}] \cdot \rho. \quad (30)$$

We should like to point out that the preceding equations (20 – 26) result from a power expansion of the expression $[1 + (c_0/c_1) \cdot O_0 \cdot O_1^{-1}]^{-1}$ in terms of a Lee series in order to resolve equation (30) with regard to $\rho$. However, equation (30) can immediately be transformed to an integral equation by the principles elaborated above, i.e. the left-hand side implies a deconvolution term of the operator $O_0$ applied to $\varphi$, whereas the operator $O_0 \cdot O_1^{-1}$ implies a convolution term $K_f$ ($s_1 > s_0$) applied to $\rho$:

$$\left. \begin{array}{l} f(\vec{x}) = \rho(\vec{x}) + \frac{c_1}{c_0} \cdot \int \rho(\vec{u}) \cdot K_f(\sigma, \vec{u} - \vec{x}) d^3u \\ \sigma^2 = s_1^2 - s_0^2 \end{array} \right\} . \quad (31)$$

With the substitution $\lambda = -(c_1/c_0)$ equation (31) represents the usual notation of an IFIE2; the inhomogeneous term $f$ results from a deconvolution procedure and $K_f(\sigma, u - x)$ is a normalized Gaussian kernel with regard to the parameter $\sigma$ in equation (31). The inverse problem is solved by finding the solution of equation (31), which can be done best with the help of LNS, i. e. the iterated kernel $K_{f(n)}$ has to be determined from the above kernel $K_f(\sigma, u\text{-}x)$. The $n^{th}$ – iterated kernel is calculated by the procedure:

$$K_{f_{(n)}}(\vec{u} - \vec{x}) = \int\int........\int K_f(\sigma, \vec{u} - \vec{u}_1) \cdot K_f(\sigma, \vec{u}_1 - \vec{u}_2) \cdot ..... \cdot K_f(\sigma, \vec{u}_{n-1} - \vec{x}) d^3u_1 d^3u_2 ....d^3u_{n-1} \quad .(31)$$

The resolving kernel $K_{res}$ is given by:

$$\left. \begin{array}{l} K_{res}(\vec{u} - \vec{x}, \lambda) = \sum_{n=0}^{L} \lambda^n \cdot K_{f_{(n+1)}}(\vec{u} - \vec{x}) \\ L \to \infty \end{array} \right\} . \quad (32)$$

The solution of the integral equation becomes:

$$\rho(\vec{x}) = \int K_{res}(\vec{u}-\vec{x},\lambda) \cdot f(\vec{u}) d^3u \quad . \quad (33)$$

With regard to practical applications we have to be aware of a finite limit L in equation (32), and L → ∞ cannot be carried out. The evaluation of the iterated terms $K_{f(n)}$ is rather simple, since $K_f$ is the normalized Gaussian kernel. Thus $K_{f(1)}$ is the normalized Gaussian kernel itself. $K_{f(2)}$ results from a composite convolution:

$$K_{f(h)} = \int K_f(\sigma,\vec{u}_1-\vec{x}) \cdot K_f(\tau,\vec{u}_1-\vec{u}) d^3u_1 = \frac{1}{\sqrt{\pi}^3} \cdot \frac{1}{\sqrt{(\sigma^2+\tau^2)^3}} \cdot \exp(-(\vec{u}-\vec{x})^2/(\sigma^2+\tau^2)) \quad (34)$$

$$\text{if} \quad \sigma^2 = \tau^2: \quad K_{f(2)} = \frac{1}{\sqrt{\pi}^3} \cdot \frac{1}{(\sqrt{2\sigma^2})^3} \cdot \exp(-(u-x)^2/2\sigma^2). \quad (35)$$

In equation (34) we have introduced the 'helping formula' $K_{f(h)}$, which allows us to determine $K_{f(3)}$, $K_{f(4)}$..., etc. by applying equation (34) iteratively. Thus by the fixation $\tau^2 = 2\sigma^2$ we obtain via equation (34) $K_{f(3)} = K_f(3\sigma^2)$. In the same fashion $K_{f(4)}$ is determined by $K_f(4\sigma^2)$ and $K_{f(n)}$ by $K_f(n\sigma^2)$. $K_{f(n)}$ appears in every order of the calculation procedure with the help *LNS*.

As already mentioned, rapid convergence is reached, if $c_0 \gg c_1$ and the ratio $\lambda$ is small. Then the powers of $\lambda$ become correspondingly much smaller. Thus for $c_0 = 0.9$ and $c_1 = 0.1$ we obtain $\lambda = -0.11111$ ($\lambda^2 = 0.01234$), whereas for $c_0 = 0.55$ and $c_1 = 0.45$ we obtain $\lambda = -0.8181$ and $\lambda^2 = 0.66942$. There is also a principal difference between the two calculation procedures with regard to the parameters $s_0$ and $s_1$. The application of the LNS method only requires $\sigma^2 > 0$, i.e. $s_1^2 > s_0^2$, while in the already presented expansion the first convolution term only exists, if $s_1^2 > 2s_0^2$ (see e.g. equation (26) for n =1, j = 0). A further difference between the two methods refers to the inverse kernel $K_g^{-1}$, which has to be determined in the first method to calculate the source function ρ from a given image function φ, whereas via LNS method we can directly calculate the source function $\rho$ from a given image function $\varphi$ without determination of the inverse kernel. The extension to a linear combination of three Gaussian convolution kernels leads with regard to the inverse problem to the following IFIE2, which assumes the shape:

$$\left.\begin{array}{l} f(\vec{x}) = \rho(\vec{x}) + \frac{c_1}{c_0} \cdot \int \rho(\vec{u}) \cdot K(\sigma_1,\vec{u}-\vec{x}) d^3u + \frac{c_2}{c_0} \cdot \int \rho(\vec{u}) \cdot K(\sigma_2,\vec{u}-\vec{x}) d^3u \\ \sigma_1^2 = s_1^2 - s_0^2; \quad \sigma_2^2 = s_2^2 - s_0^2 \end{array}\right\} . (36)$$

In order to evaluate equation (36) by equation (31), we write this equation in the form:

$$\left.\begin{array}{l} f(x) = \rho(x) - \lambda \cdot \int \rho(u) \cdot [K(\sigma_1,u-x) + \alpha \cdot K(\sigma_2,u-x)] du \\ \lambda = -c_1/c_0; \quad \alpha = c_2/(c_0 \cdot c_1) \end{array}\right\} . (37)$$

For the evaluation of the inverse kernel we need to calculate $K_{f(n)}$:

$$K_{f_{(n)}} = \int\int \cdots \int [K_f(\sigma_1, \vec{u}-\vec{u}_1) + \alpha \cdot K_f(\sigma_2, \vec{u}-\vec{u}_1)] \cdot [K_f(\sigma_1, \vec{u}_1-\vec{u}_2) + \alpha \cdot K_f(\sigma_2, \vec{u}_1-\vec{u}_2)] \cdot \\ \cdots \cdot [K_f(\sigma_1, \vec{u}_{n-1}-\vec{x}) + \alpha \cdot K_f(\sigma_2, \vec{u}_{n-1}-\vec{x})] d^3u_1 d^3u_2 \cdots d^3u_{n-1} \quad .(38)$$

It is evident that $K_{f(n)}$ has to contain the terms $K_f(n \cdot \sigma_1, u-x)$ and $\alpha^n \cdot K_f(n \cdot \sigma_2, u-x)$, but the binominal theorem also provides mixed products, and by evaluation of equation (38) $K_{f(n)}$ assumes the shape:

$$K_{f_{(n)}}(\vec{u}-\vec{x}) = \sum_{j=0}^{n} \alpha^j \cdot \binom{n}{j} \cdot K_f((n-j)\cdot\sigma_1^2 + j\cdot\sigma_2^2, \vec{u}-\vec{x}). \qquad (39)$$

It must be pointed out that now equation (32) has to be evaluated with the help of equation (39). With regard to convergence aspects in the above cases according to equations (35 – 39) it is obvious that convergence is fast, if $c_0$ satisfies $c_0 \gg c_1$ or $c_0 \gg c_2$, i.e. the leading term refers to $c_0$ and the additional contributions only represent (small) long-range tails. Please note that the LNS method is also applicable, if $c_1 < 0$ ($c_0 + c_1 + c_2 = 1$) is assumed. Examples for this case will be presented in the following section.

*2.3. Monte-Carlo methods*

With regard to problems of image processing (e.g. blurred images due to scatter effects) we have carried out Monte-Carlo calculations with the EGSnrc code [28]. This code has been applied most widely to various tasks in medical radiation physics. We have performed Monte Carlo calculations using the EGSnrc code with regard to problems of image processing in the *MV- and KV-domain*.

*2.3.1. Image processing in* **the** *MV-domain*

Absorption – and attenuation curves, transverse profiles in various depths for the simulation of radiation responses of a detector array (portal imaging) have been determined for field sizes 1 x 1 cm$^2$ up to 20 x 20 cm$^2$. Previous results have been used with regard to the energy spectrum of 6 MV [4].

*2.3.2. Image processing in the KV-domain*

We have determined the energy spectrum of 100 KV and 125 KV photons of CT/CBCT and the absorption/scatter behavior in some media of relevance, e.g. water-equivalent and phantoms with different material densities (lung, bone). A main purpose was the connection between Hounsfield units and the scatter parameters required for the 2D scatter kernel:

$$K_g = c_0 \cdot K_0(s_0, u-x, v-y) + c_1 \cdot K_1(s_1, u-x, v-y). \qquad (40)$$

In general, the scatter parameters $s_0$ and $s_1$ depend (increase) on the depth z, and this is the way to treat the depth-dependent scatter of a pencil beam. The correspondence between the Hounsfield value and electron density ρ is well-established. The scaling of the scatter parameters $s_0$ and $s_1$ can be scaled according to the electron density ρ, if the scatter parameters are known for water. A possible, but rather intricate way to eliminate scatter in CT/CBCT images would be obtained by the deconvolution of photon pencil beams, i.e. the methods of radiation therapy planning [4] are transformed to image processing.

Therefore, we extend here a previously developed method of the deconvolution according to the volume [2] to the parallel solution procedure of LNS presented in this study.

### 2.3.3. Extension of the LNS method to volume-dependent scatter functions s, $s_0$, $s_1$ and $s_2$

We have already pointed out that the scatter parameters s, $s_0$, $s_1$ (and eventually $s_2$) have by no means to be constant values. Thus, in the pencil beam algorithms [3 – 4] these parameters are not constant, but they represent scatter functions depending on the depth z. However, this restriction is, in general, not necessary in all formulas we have developed in this study:

The differential operator formulations of one and/or more than one kernel expressed by $O^{-1}$, $O$, $O_g^{-1}$, $O_g$, permit a dependence of all parameters s, $s_0$, $s_1$, $s_2$ and related composite terms like σ, $σ_1$, $σ_2$ of all three dimension magnitudes x, y, z, since the differential operators in the exponential functions are not influenced by this property. This property is also true with regard to all integral operator formulations (inclusive IFIE2 and LNS procedure), where the half-width parameters do not affect the integration variables. In all our applications, we do not account for neither complex-valued Gaussian kernel functions nor source/image functions ρ and φ. We restrict ourselves to positively definite source/image functions. Thus we have previously used Fourier expansions of the scatter functions [2], and the same procedures are now applied to some cases of the LNS calculations (image problems):

$$s_k(x,y,z) = \sum_{j1,j2,j3=0}^{\infty} A^k_{j1,j2,j3} \cdot \cos(\vec{j}\cdot\vec{x}) + B^k_{j1,j2,j3}\cdot\sin(\vec{j}\cdot\vec{x}) \bigg\} \quad (41)$$
$$k = 0, 1, 2$$

Equation (41) is of particular importance, if the source function ρ is connected to a dose distribution function D without scatter, i.e. it only contains absorption but not attenuation, whereas the image function φ represents a blurred dose distribution which also contains scatter.

### 2.4. Measurement data and calculations via therapy planning system

In this communication we have used the algorithm AAA implemented in the planning system Eclipse[R] (Varian, installation in the Klinikum Frankfurt/Oder). This algorithm has been previously published [4]. The radiation leaving a phantom has been recorded with the Iview[R] (Synergy, Elekta). Details of CT/CBCT measurements have been previously given [2].

## 3. Results

Since we are mainly interested in calculation results and the reliability obtained by the LNS expansion, the following section we account for those examples we have already discussed in detail by published methods [2]. By that, the whole algorithm concerning the inverse problem of linear combinations of Gaussian convolution kernels will obtain more flexibility.

*3.1. Comparison of LNS procedure with a previously published method*

In order to check the reliability and convergence properties of the LNS procedure, we perform at first applications we have previously obtained by the calculation of $K_g^{-1}$. In both calculation procedures, we need the deconvolution kernel $K_0^{-1}(s_0, u-x)$, which has to be appropriately extended, if necessary, to more than one spatial dimension. Since $K_0^{-1}$ represents itself an infinite expansion, we denote here the finite break off value by *N* we have used in a corresponding calculation. It has to be pointed out that for a reasonable comparison of the two different inverse procedures N has to be identical in both cases. The finite break off value of the sequence of Gaussian convolution terms according to equations (26 - 27) will be denoted by *M*, and the related value of the LNS procedure according to equations (26, 31 - 33) by *L*. The best test of the derived deconvolution formulas can be obtained by corresponding convolutions of some model cases and back calculations via LNS procedure. Since the deconvolution formulas represent order-by-order calculations, a principal aim of the tests was to specify the required order and precision to obtain the source function (origin) in a satisfactory way.

The principal problems of deconvolutions and possible pitfalls can be verified either by the figures 1 – 5 or by figures 12 - 14 in section 3.2. These figures show that rather different sources (e.g. three adjacent boxes or boxes with an empty space between them) with different *rms* values s, $s_0$, $s_1$ and $s_2$ lead to similar image functions. By that, we have to verify that the underlying *rms* values have to be known rather exact from measurement data or by Monte-Carlo calculations to prevent artifacts by the deconvolution procedures. Only due to the very accurate knowledge of the subjected convolution parameters it is possible that the inverse procedure also reliable works with sufficient accuracy. The so-called 'try-and-error' method with certain start values for the *rms* parameters s, $s_0$, $s_1$, $s_2$ and coefficients $c_0$, $c_1$, $c_2$ might lead to artifacts. The model cases according to figures 1 – 5 and 12 – 14 may also have a practical importance, since the boxes represent finite step functions, where the $L_1$-integrability is rather favorable to handle, and the deconvolution via Fourier transforms and Wiener filters leads to diverging jumps at the edges (this is a typical ill-posed problem [22]). In radiotherapy the fluence determination and optimization within finite intervals (grid size) and modulation represents an important feature in IMRT (or Rapid Arc) therapy. On the other side, only by rescaling of the underlying geometry we are directly guided to these aspects and novel treatment schemes of modern radiotherapy [20], which appear to lead to a better protection of critical organs and to fulfill the corresponding constraints of radiobiology and radiation protection.

1414

Table 1. Convolution/deconvolution parameters of figures 1 and 2.

| Figure | $c_0$ | $c_1$ | $s_0$/cm | $s_1$/cm | L | M | N |
|---|---|---|---|---|---|---|---|
| 1/dashes | 1 | - | 0.1 | - | - | - | 4 |
| 1/dots | 0.8 | 0.2 | 0.025 | 0.075 | 5 | 4 | 4 |
| 2/dashes | 1 | - | 0.1 | - | - | - | 7 |
| 2/solid | 0.8 | 0.2 | 0.025 | 0.075 | 8 | 7 | 7 |

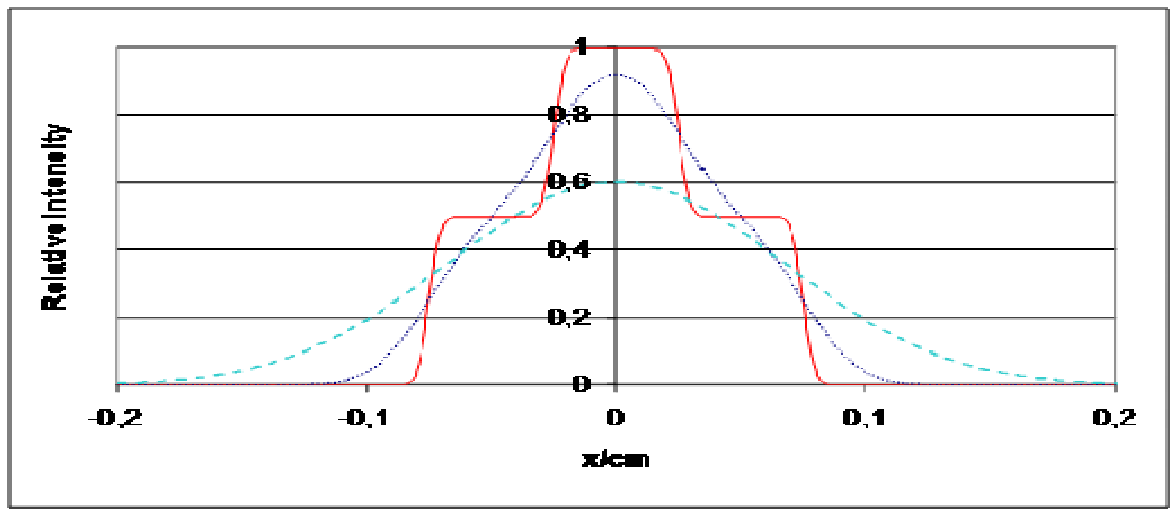

**Figure 1.** Source profile is based on the geometry: three adjacent boxes with box width of 0.05 cm. Height of the central (middle) box is 1; the height of the adjacent boxes (left- and right-hand side) is 0.5. Dashes: convolution with one Gaussian kernel. Solid line: deconvolution containing rounded corners. The precision of these deconvolutions is not very high yielding an increased roundness at the edges.

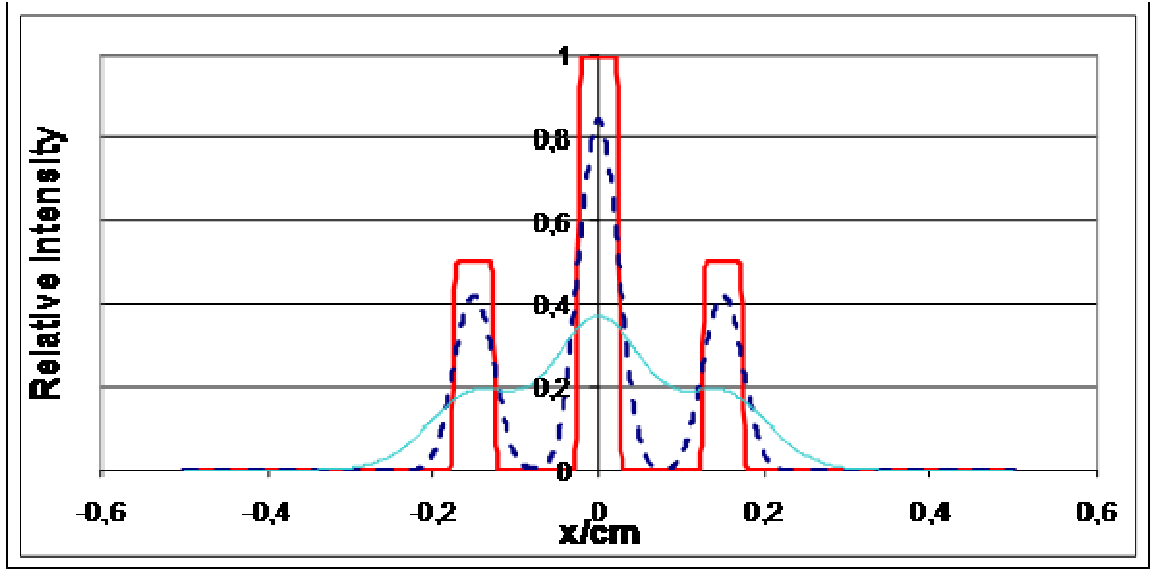

**Figure 2.** Three boxes (peaks) with box width 0.05 cm and 0.1 cm space between the boxes. Solid line: one Gaussian convolution; dashes: two kernels. Solid line (bold): deconvolution containing small rounded corners. The desired precision has been increased.

**Table 2.** Convolution/deconvolution parameters of figures 3 – 6.

| Figures | $c_0$ | $c_1$ | $c_2$ | $s_0$/cm | $s_1$/cm | $s_2$/cm | L | M | N |
|---------|-------|-------|-------|----------|----------|----------|---|---|---|
| 3 – 4   | 0.90  | -0.38 | 0.48  | 0.40     | 0.82     | 1.50     | 11 | *10* | *10* |
| 5 – 6   | 0.90  | -0.38 | 0.48  | 0.44     | 0.78     | 1.55     | 10 | 11 | 10 |

There is a principal difference between the parameters stated in Table 1 and Table 2. The parameters of table 2 refer to that case, where we have accounted for a linear combination of three Gaussian convolution kernels, but with $c_0$ and $c_2 > 0$ and $c_1 < 0$ ($c_0 + c_1 + c_2 = 1$). This case is also supported by the LNS procedure and increases the flexibility of convolution/deconvolution applications without having to consider the Mexican hat problem with $c_1 < 0$ and $c_2 = 0$. We should recall that in spite of the modification with $c_1 < 0$ the condition $K_g \geq 0$ has to be satisfied. This property certainly represents a constraint for the choice of $c_1$ and $s_1$.

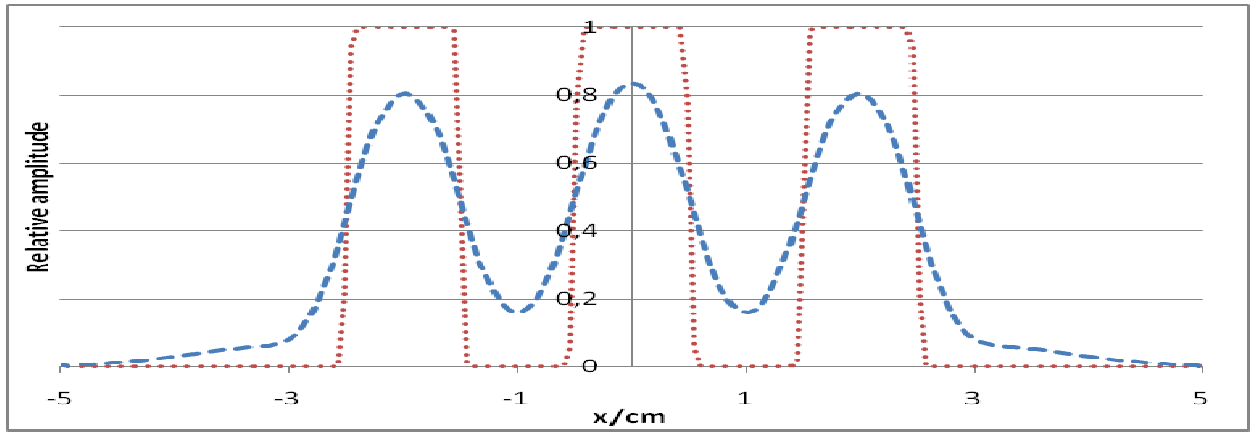

**Figure 3.** Convolution and deconvolution of three boxes (box length: 1 cm, space between them: 1 cm).

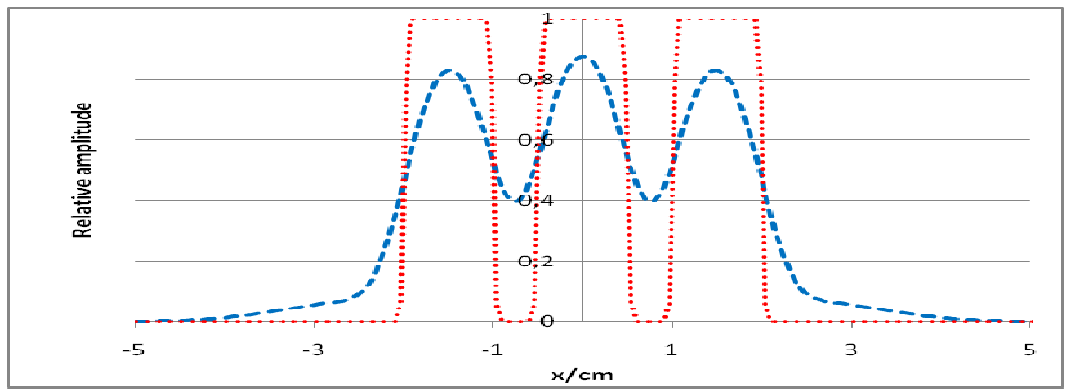

**Figure 4.** The content of this figure is equivalent to figure 3, only the space between the boxes is reduced to 0.5 cm.

In contrast to figure 4 the distance between the three boxes is increased in figure 5; the relative amplitude between the boxes obtained after convolution reflects this property.

This property represents an essential restriction with regard to the choice of the scatter parameters and coefficients of the linear combinations $c_0$, $c_1$ and $c_2$. The negative value of $c_1$ yields the rapid decrease of the relative amplitude at the outermost boxes. The relative amplitude is not yet specified; it might refer to a fluence or dose distribution or to a signal strength in some other kinds of applications, where convolutions and their inverse problems are applied (e.g. image processing based on magnetic resonance tomography).

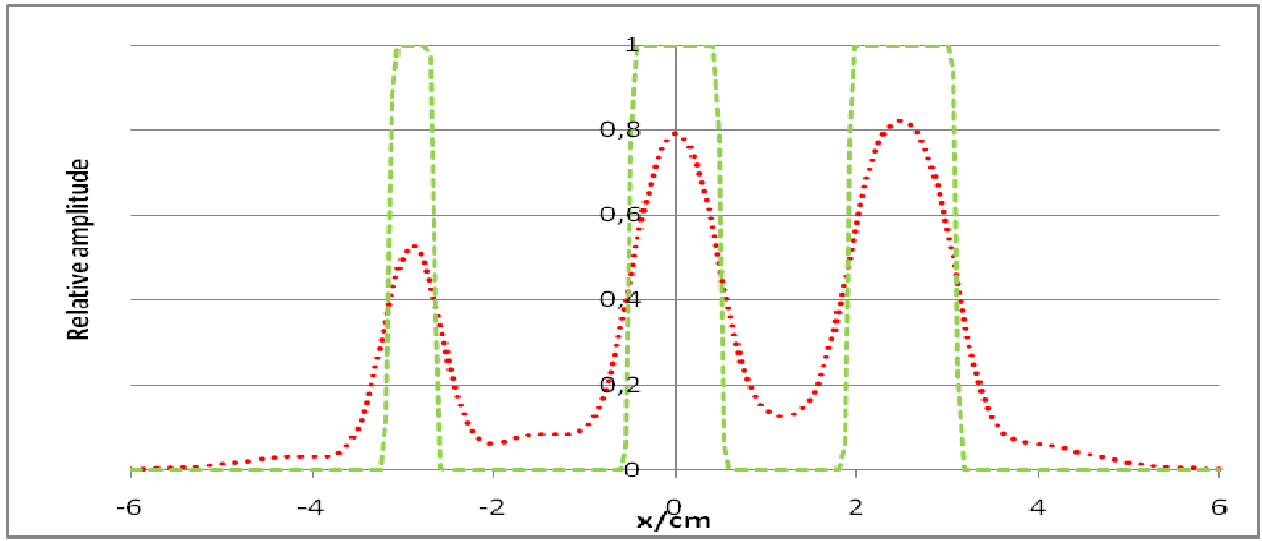

**Figure 5.** Convolution/deconvolution of three boxes (box lengths: 1 cm (right-hand side), 0.8 cm (central part), 0.5 cm (left-hand side); spaces: 1.6 cm (right-hand side), 2.35 cm (left-hand side)).

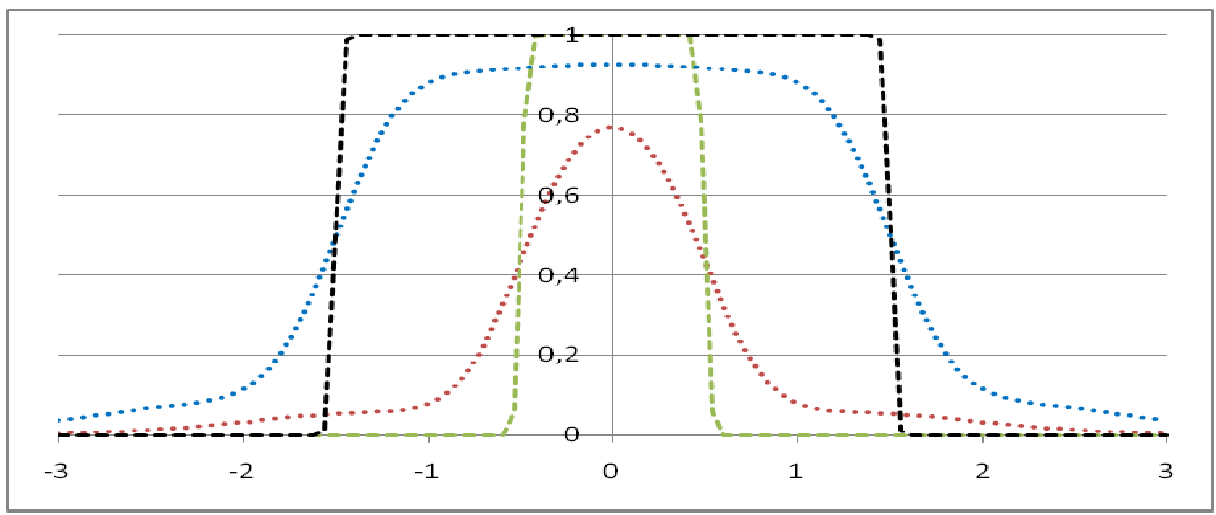

**Figure 6.** Convolution (dots) and deconvolution (dashes) of two boxes (box lengths: 1 cm and 2 cm).

The following examples (figures 7 – 11) represent a modification of a previously considered image deconvolution [2] of a test phantom (CT image), whereas we now consider the same phantom configuration in connection with a CBCT image. The test phantom (figure 7) consists of an inner cylinder with a diameter of 4 cm (HU = 700) embedded by an outer tube containing water-equivalent material (HU = 0); the total phantom diameter amounts to 16 cm. The impinging photon beam with 140 KV now is a broad beam (CBCT), which can be calculated from a Gaussian beamlet with $s_0 = 0.87$ mm, whereas in the previous study we have used a Gaussian beamlet with 125 KV (scanning technique with CT, $s_0 = 0.5$ mm).

Please note that the removal of noise of detectors recording images ('raw data') has been performed by a smoothing function established in the algorithm (figures 7 – 10 and 15 – 16). The application of a deconvolution procedure as previously used [2] or in this study via LNS requires the determination of the scatter in the phantom. The energy spectrum of the incident photon beam has been determined by Monte-Carlo calculations [28]. In both cases, CT and CBCT, a detector array records the attenuation radiation behind the phantom. The valid scatter functions for the CT imaging have already been presented [2], and the proportionality between the electron density $\rho_{el}(x, y, z)$ and the scatter functions $s_0(x, y, z)$, $s_1(x, y, z)$, and $s_2(x, y, z)$ holds ($c_0$, $c_1$, $c_2$ remain unchanged). However, the end values of the scatter functions at the detector plane are not valid. The scaling transformation has to be corrected by the detector influence and the initial scatter of the photon beam at the impinging position:

$$\left.\begin{array}{l} s_0(x, y, z) = s_{0,i} \cdot \rho_{el}(x, y, z) \\ s_1(x, y, z) = s_{1i} \cdot \rho_{el}(x, y, z) \\ s_2(x, y, z) = s_{2,i} \cdot \rho_{el}(x, y, z) \\ s_{0i} = 0.42 \ cm; \ s_{1i} = 0.56; \ s_{2i} = 2.72 \ cm \\ c_0 = 0.87; \ c_1 = -0.12; \ c_2 = 0.25 \end{array}\right\} . \qquad (42)$$

The number of linear combinations of kernels (two kernels for CT and three kernels for CBCT) is the principle difference between the parameters according equation (42) valid for CBCT and those parameters valid for CT. Moreover, $s_{1i}$ does not satisfy $s_{1i} > \sqrt{2} \cdot s_{0i}$; therefore the use of the previous algorithm is difficult to handle; a detailed treatment of this situation has been previously given [2]. The uncorrected electron density functions result from the Fourier expansion (41), where the scatter influence is accounted for. We have now to perform the task that the deconvolution of the 3D image at the central ray should provide the same result as one image of CT.

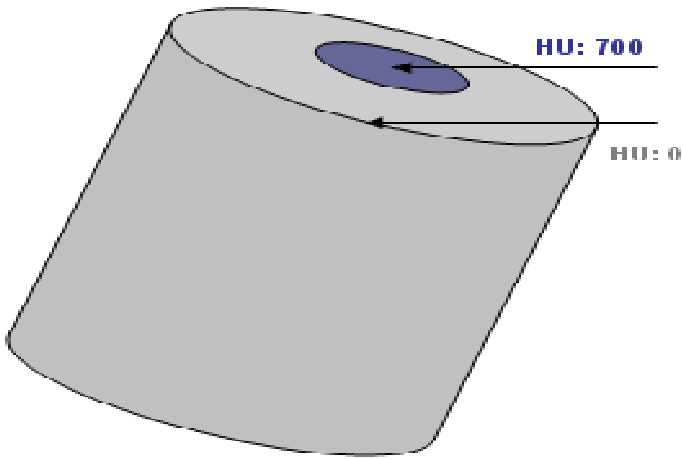

**Figure 7.** Phantom: water-equivalent material/bone. Inner cylinder: bone with HU = 700, outer part: water-equivalent material with HU = 0. In a later application the bone material with HU = 700 will be replaced by air and serves as a test phantom for a portal imager of a linear accelerator.

In the case of CT image processing, the cylinder is scanned along the cylinder axis, whereas in CBCT image processing the image is produced via one rotation by divergent broad beam. The problem of scanning by CBCT is certainly an increased contribution of scatter by the X-rays. This is a characteristic feature of all cases of broad beams and not only restricted to the KV domain.

Figure 8 refers to the convolution/deconvolution problem of the test phantom obtained by CT. In the CBCT case the situation is more difficult because of the divergence of the broad beam, and Figure 9 presents the convolution/deconvolution of the central ray (i.e. without divergence).

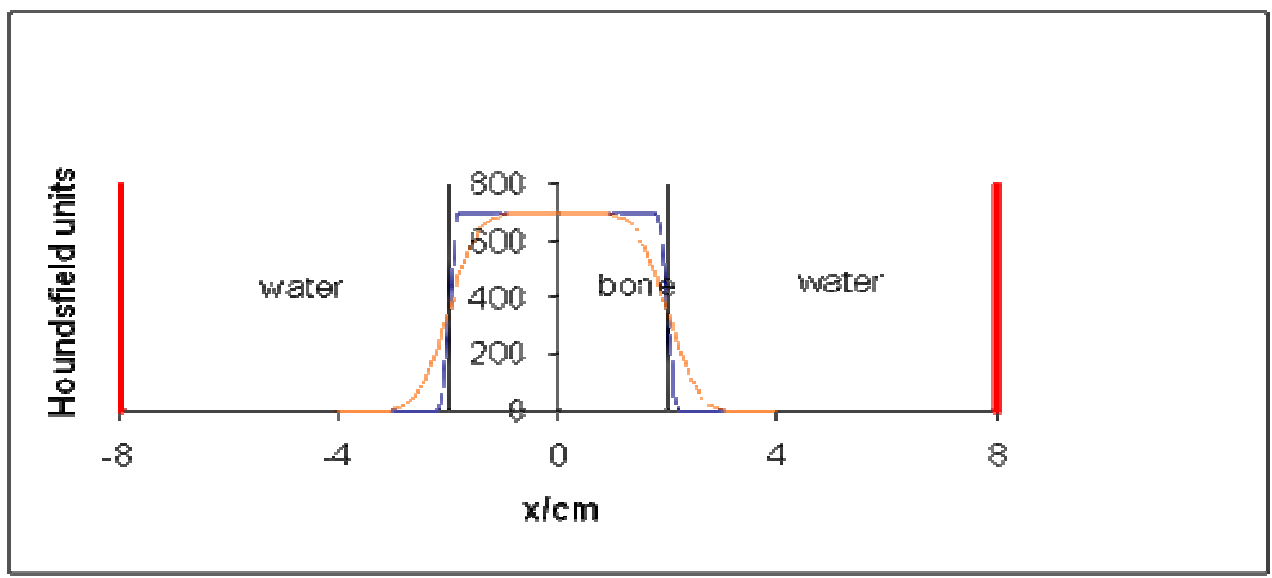

**Figure 8.** Profile of the Hounsfield units (CT) of the phantom cylinder (N = 4, M = 4, L = 5).

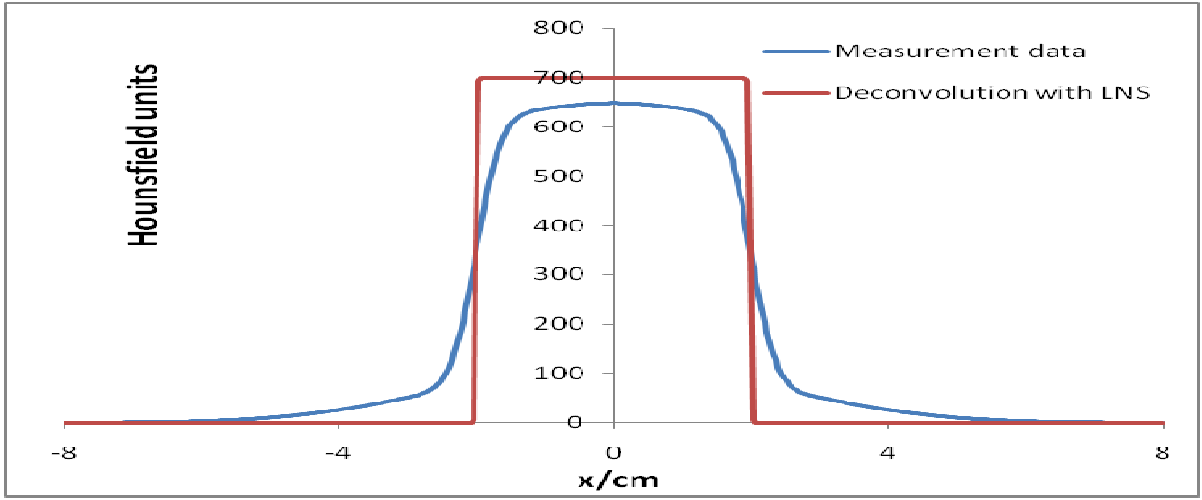

**Figure 9.** Profile of the Hounsfield units (CBCT) of the phantom cylinder (N = 7, L = 8). Figure 9 represents the CBCT correspondence of Figure 8. The Hounsfield unit of water amounts to 0.

Equation (42), which determines the space-depending scatter function useful for deconvolutions of the complete volume, has to be modified in the CBCT case:

$$\left.\begin{array}{l} s_0(x,y,z) = s_{0,i} \cdot \rho_{el}(x,y,z) \cdot Cf \\ s_1(x,y,z) = s_{1,i} \cdot \rho_{el}(x,y,z) \cdot Cf \\ s_2(x,y,z) = s_{2,i} \cdot \rho_{el}(x,y,z) \cdot Cf \end{array}\right\}. \tag{43}$$

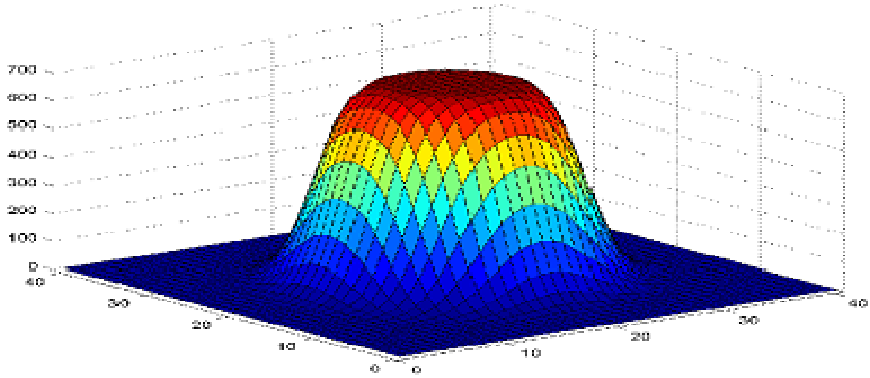

**Figure 10.** Hounsfield units of a 2D cylinder based on a measurement with a detector array.

The correction factor $Cf$ results from the divergence of the X-rays. Only for the central ray we have to put $Cf = 1$.

We have to point out that it is a feature of CT that divergent rays are not used. However, in our case of CBCT application with rotational symmetry the factor $Cf$ is determined by

$$Cf = \sqrt{(d^2 + SAD^2)/SAD^2} \ . \qquad (44)$$

In equation (44) SAD refers to the source-axis-distance and d to distance from the center of the central axis of the cylinder and its rotation axis.

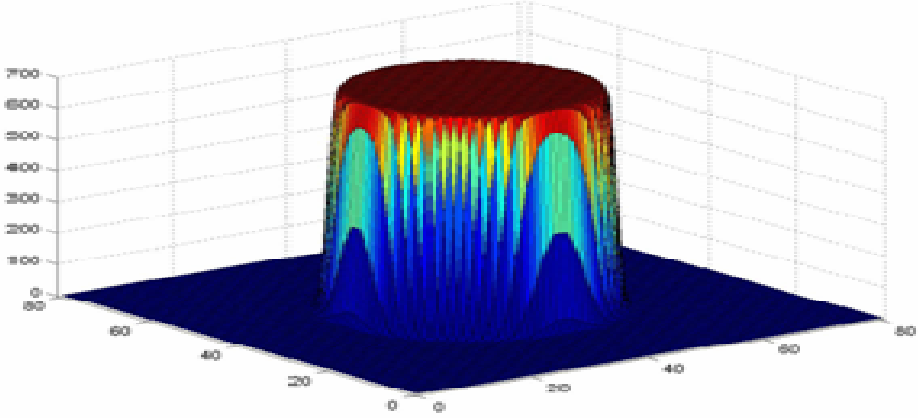

**Figure 11.** Hounsfield units of a 2D cylinder (Figures 8 - 9 represent the result after deconvolutions).

Thus figure 11 is chosen such that the central ray of CBCT scanning is identical with the result of CT scanning. It is obvious that we need significantly more effort in the CBCT case with regard to the inverse problem, namely the order L of the deconvolution procedure. On the other side, this scanning technique provides a complete 3D image. In order to obtain reliable results in the CBCT case, the calculations had to be performed by accounting for higher order terms in the LNS procedure. In this communication, we have only considered the inverse problem of the central ray, but with regard to CBCT the calculation procedure of the inverse problem requires the modifications according to equation (44).

*3.2. Further results obtained by LNS*

The examples presented in figures 11 – 13 may serve as further tests of inverse calculations via LNS with possible applications to IMRT/IGRT. A comparison of figure 12 with figures 13 – 14 demonstrates the possible pitfalls of deconvolutions. Thus it is clear that the convolution of a triangle provides a 'triangle' with rounded corners. However, the shape of the images obtained via convolution of non-adjacent boxes might lead to the assumption that the source function has also the shape of a triangle, which is apparently not true. This fact clearly demonstrates that 'try-and-error' methods to determine the parameters for the inverse procedures (e.g. LNS method) might lead to artifacts.

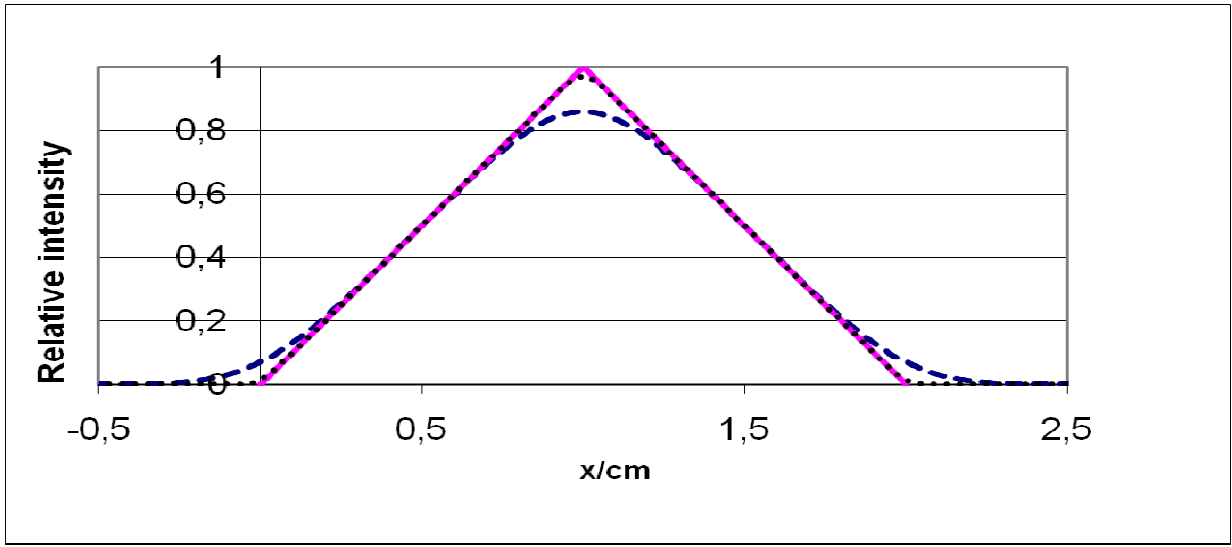

**Figure 12.** Convolution of a triangle (solid) with one Gaussian kernel (dashes) and deconvolution (dots).

**Table 3.** Convolution/deconvolution parameters in figures 11 – 13.

| Figure | $c_0$ | $c_1$ | $s_0$/cm | $s_1$/cm | L | M | N |
|---|---|---|---|---|---|---|---|
| 12 | 1 | - | 0.25 | - | - | - | 4 |
| 13/dashes | 1 | - | 0.10 | - | - | - | 12 |
| 13/dots | 0.80 | 0.20 | 0.025 | 0.075 | 15 | 15 | 12 |
| 14/solid | 1 | - | 0.09 | - | - | - | 12 |
| 14/dashes | 0.80 | 0.20 | 0.015 | 0.050 | 15 | 16 | 12 |

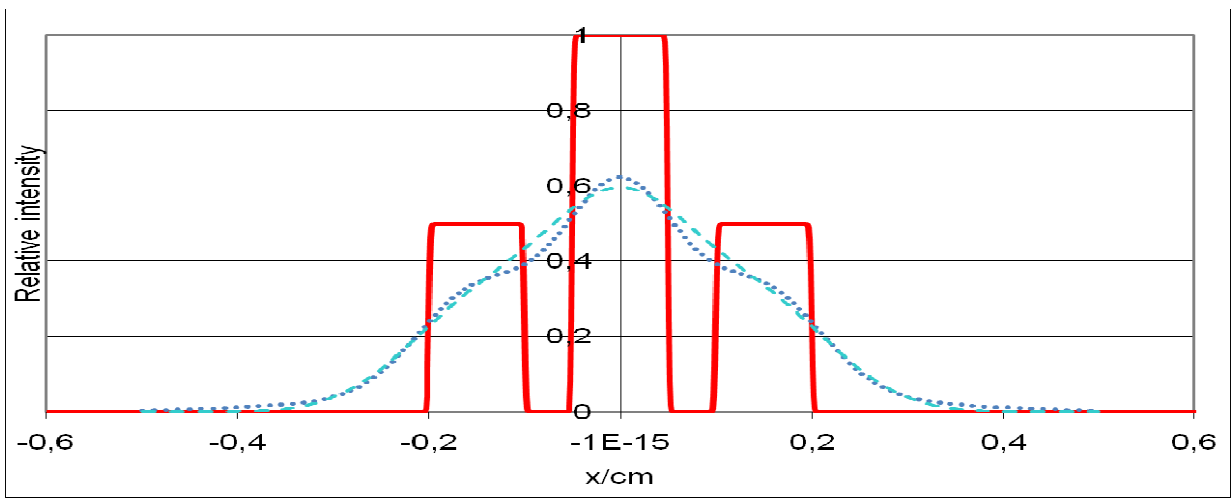

**Figure 13.** Convolution of three boxes (box length: 0.1 cm, space length between the boxes: 0.05 cm, height of the source functions: 1 (middle part) and 0.5 (at both sides)). Deconvolution: identical with the solid boxes, rounded corners not verifiable.

The deconvolution procedure by the LNS method has been applied (L = 20, N = 20) in figure 15. The reason for the increased effort results from the long-range tail of the scatter of the high energy bremsstrahlung.

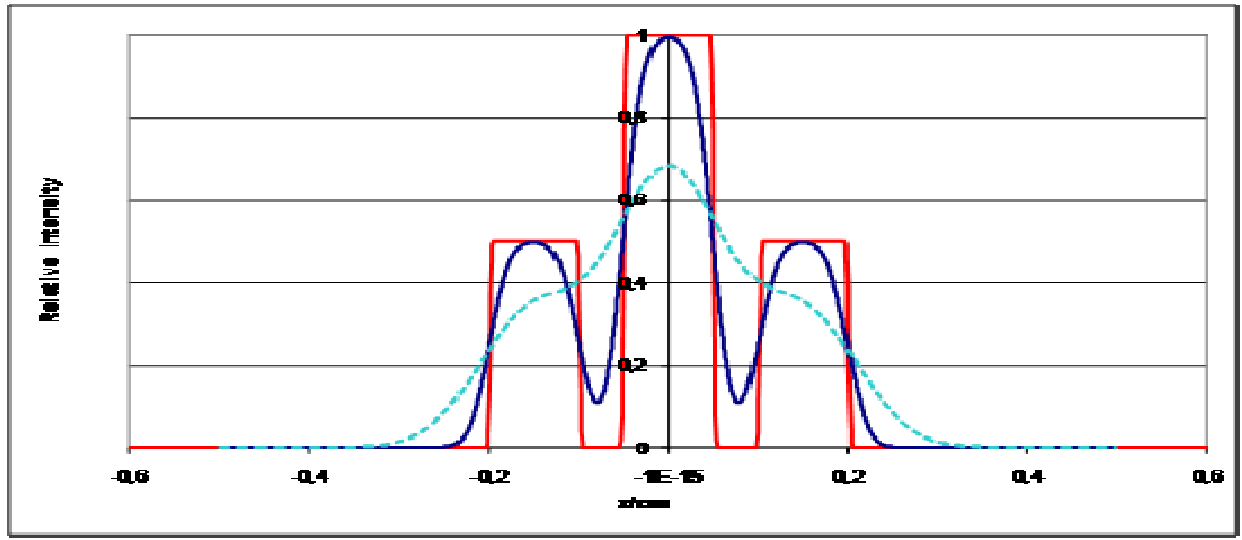

**Figure 14.** Geometry and box heights: see figure 12. Convolutions have been obtained with different parameters. Deconvolutions are considered as identical with the origin (solid boxes), if the rounded corners are not verifiable.

The final application presented in this study refers again to figures 7 – 11. Instead of image creation with X-rays (CT, CBCT) a portal imager (6 MV, bremsstrahlung) has been applied. Due to the long range of lateral scatter the portal imager does not provide the same height of the central ray as is can be verified from the previous figures (KV domain), and the lateral tail has significantly been increased. The deconvolution procedure has also to be performed by accounting much more terms of higher order in the LNS procedure than in the previous cases, and some noteworthy roundness can be verified in spite of the increased effort with regard to the order L.

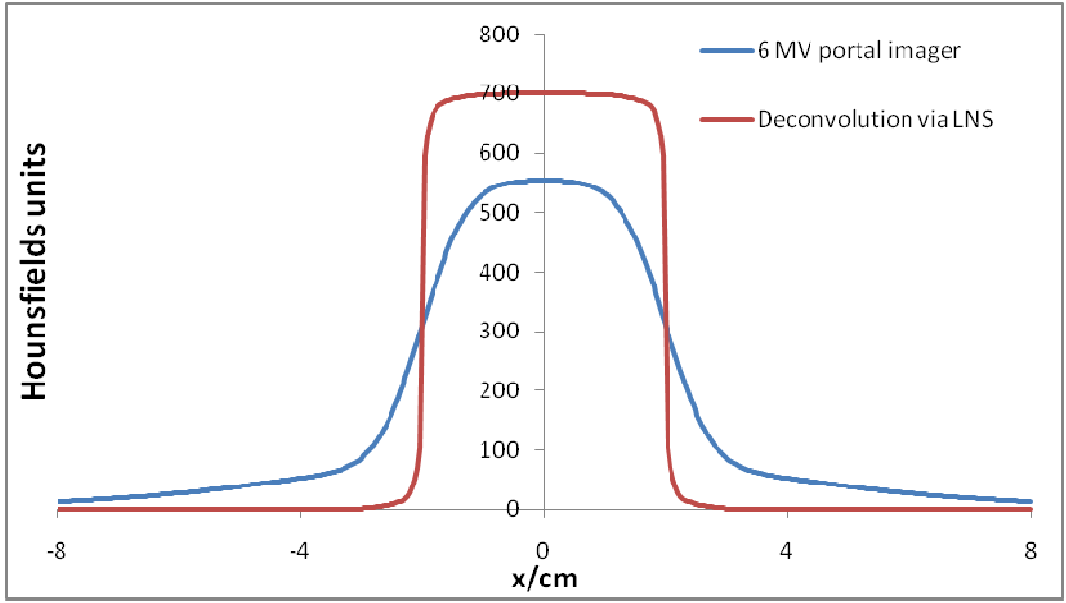

**Figure 15.** Water/bone phantom (figure 7) and image produced by a portal imager, 6 MV, bremsstrahlung).

The *rms* parameters for the application of the LNS method have been used from a previous publication [4]:

$$\left.\begin{array}{l} c_0 = 0.66;\ c_1 = 0.26;\ c_2 = 0.08 \\ s_0 = 0.82\ cm;\ s_1 = 4.735\ cm;\ s_2 = 12.334\ cm \end{array}\right\}. \quad (45)$$

The calculation with the previous method (M = 29, N = 20, [2]) showed a superiority of the LNS procedure to provide faster convergence in the case of long-rang tails.

We have also used a modified configuration of the cylinder according to figure 7, namely with air in the inner part instead of bone-equivalent material, for a further measurement with a portal imager (6 MV, bremsstrahlung). Based on the LNS method the measurement data have been analyzed with parameters of the publication cited above [4]:

$$\left.\begin{array}{l} c_0 = 0.66;\ c_1 = 0.26;\ c_2 = 0.08 \\ s_0 = 0.39\ cm;\ s_1 = 1.77\ cm;\ s_2 = 6.16\ cm \end{array}\right\}. \quad (46)$$

Thus figure 16 shows the adaptation of the measurement data with the help of the AAA algorithm and the deconvolution via LNS (boxes with weak roundness at the corners). In contrast to figure 15, where we have used Hounsfield units as the reference scale, we present in figure 16 the density (cross-section).

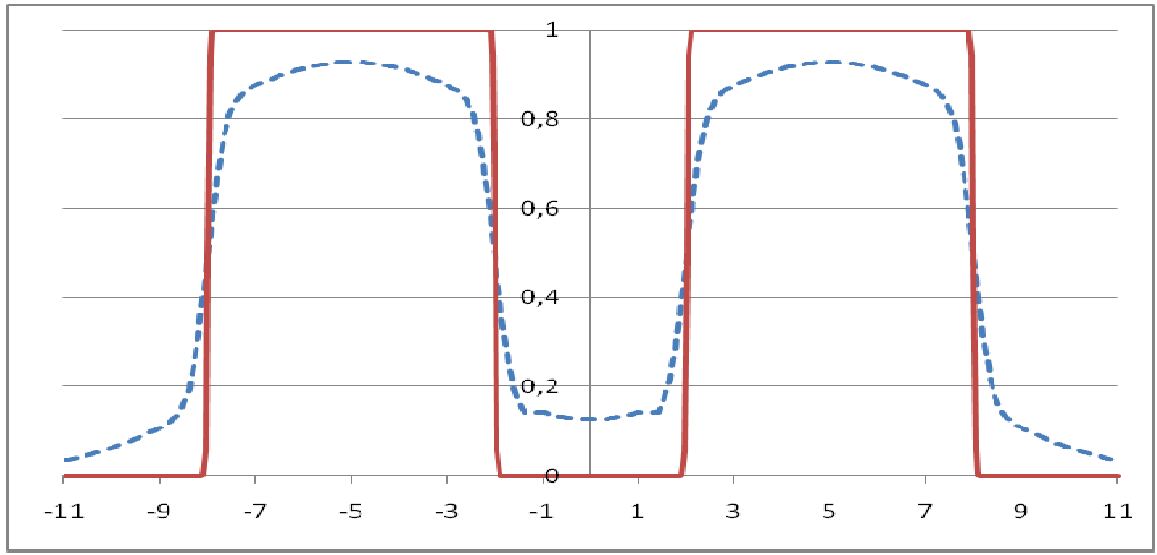

**Figure 16.** Water/air phantom as a modified configuration of figure 7 and image produced by 6 MV bremsstrahlung in a portal imager system (L = 11, N = 10).

Figures 15 and 16 indicate the superiority of LNS with regard to inverse problems in those cases, where the long-range tails are rather significant.

## 4. Conclusions

The property of scatter functions to account for their 2D or 3D dependence; this fact simplifies to determine the origin images by a formal way, i.e. the removal of the scatter via a calculation procedure. Scatter processes represent an inevitable property of imaging. Besides these aspects of the inverse problem (image processing) we are also able to mention the fluence determination in IMRT/IMPT and refer to a specific publication of this application, where the inverse problem of Gaussian convolution plays a significant role [20, 21]. The discussed model cases of adjacent and nonadjacent boxes The preceding sections show that the application of the LNS method provides an attractive alternative way to solve the inverse problem (deconvolutions) of the determination of the origin image (source functions), which have been blurred by scatter of high energy photons (KV- and MV-domain). The method can be best demonstrated by model cases (phantoms). In particular, we are able to exploit show that with regard to inverse calculations one has to be very careful in order to avoid artifacts produced by improper scatter parameters. We particular point out the problem of noise produced by detectors, which may lead to difficult decisions, whether the origin function contains real peaks or result from fluctuations of detector properties. Without profound knowledge of these parameters and further empirical experience in their handling it appears impossible to obtain reliable results of complex problems, which are confronted in CT/CBCT imaging. In order to restrict the scope of this study we have been unable to account for MR or PET imaging, although the latter two disciplines have become a very important tool in many other domains of medicine, which are rather different from radiology and radiotherapy, e.g. neurology, surgery and molecular image processing in pharmacology.

## References

[1] B. Gottschalk, A. Koehler, R. Schneider, J. Sisterson and M. Wagner, Multiple Coulomb scattering of 160 MeV protons *Nuclear Instrument Methods* **B74** (1993) 467.

[2] W. Ulmer, Inverse problem of a linear combination of Gaussian convolution kernels (deconvolution) and some applications to proton/photon dosimetry and image processing *Inverse Problems* **26 (**2010) 085002.

[3] W. Ulmer and E. Matsinos, Theoretical methods for the calculation of Bragg curves and 3D distribution of proton beams *European Physics Journal (ST)* **190 (**2011) 1 - 72.

[4] W. Ulmer, J. Pyyry and W. Kaissl, A 3D photon superposition/convolution algorithm and its foundation on results of Monte-Carlo calculations *Physics in Medicine and Biology* **50** (2005) 1767.

[5] W. Ulmer and B. Schaffner, Foundation of an analytical proton beamlet model for inclusion in a general proton dose calculation system *Radiation physics and chemistry* **80 (**2011) 378.

[6] W. Ulmer and W. Kaissl, The inverse problem of a Gaussian convolution and its application to the finite size of the measurement chambers/detectors in photon and proton dosimetry *Phys. Med. Biol.* **48** (2003) 707.

[7] F. Crop, N. Reynaert, G. Pittomvils, L. Paelinck, C. De Wagner, L. Vakaet and H. Thierens, The influence of small field sizes, penumbra, spot sizes and measurement depth on perturbation factors for microionization chambers *Phys. Med. Biol.* **54 (**2009) 2951.

[8] I. Das, G. Ding and A. Ahnesjö, Small fields: nonequilibrium radiation dosimetry *Med. Phys.* **35 (**2008) 206-25.

[9] H. Palmans, Perturbation factors for cylindrical ionization chambers in proton beams: Part I. Corrections for gradients *Phys. Med. Biol.* **51 (**2006) 3483.

[10] E. Pappas, T. Maris, A. Papadakis, F. Zacharopoulou and J. Damilakis. Experimental determination of the effect of detector size on profile measurements in narrow photon beams *Physics in Medicine and Biology* **33** (2006) 3700.

[11] W. Ulmer, On a unified treatment of diffusion and kinetic processes *International Journal of Quantum Chemistry* **23** (1983) 1931.

[12] W. Ulmer, On a unified treatment of kinetics and diffusion, and a connection to a nonlocal Boltzmann equation *International Journal of Quantum Chemistry* **27** (1985) 203.

[13] V. Isakov, The inverse conductivity problem with limited data and applications *Journal of Physics: Conference Series* **73 (**2007) 012011 – 012020.

[14] T. Matsuura and S. Saitoh, Analytical and numerical inversion formulas in the Gaussian convolution by using the Paley-Wiener spaces *Applicable Analysis* **85** (2006) 901.

[15] S. Saitoh, Approximate real inversion formula of a Gaussian convolution *Applicable Analysis* **83** (2004) 727.

[16] G. McLachlan, T. Krishnan, *The EM Algorithm and Extensions Wiley Series in Probability and Statistics* (New: York: Wiley, 1997).

[17] M. Figueiredo and R. Nowak, An EM algorithm for wavelet-based image restoration *IEEE Trans. Image Process.* **12 (**2003) 906.

[18] T. Tuminonen, Perfusion deconvolution via EM algorithm *PhD Thesis* University of Helsinki, Research Project Applied Mathematics, 2004.

[19] T. Bortfeld, T. Chan , A. Trofimov and J. Tsitsiklis, Robust management of motion uncertainty in intensity-modulated radiation therapy *Oper. Res.* **56 (**2008) 1461.

[20] Y. Fan and R. Nath, Intensity modulation under geometrical uncertainty: a deconvolution approach to robust fluence *Phys. Med. Biol.* **5** (2010) 4029.

[21] C. Bäumer and J. Farr, Lateral dose profile characterization in scanning particle therapy *Med. Phys. 38 (*2011) 2904 – 2013.

[22] T. Ming Fang, S. She, R. Nagem and G. Sandri, Convolution and deconvolution with Gaussian kernel *Nuovo Cimento* **B 109** (1994) 83.

[23] F. Garcia-Vicente, J. Delgado and C. Rodriguez, Exact analytical solution of the convolution integral equation for a general profile fitting function and Gaussian detector kernel *Phys. Med. Biol.* **25** (2000) 202.

[24] J. Liu, J. Yao and R. Summers, Scale-based scatter correction for computer-added polyp detection in CT colonography *Medical Physics* **35** (2008) 5564.

[25] A. László, A robust iterative unfolding method for signal processing *J. Phys*. A 39 (2006) 13621 – 13640.

[26] G. Micolau, J. Postel-Pellerin, R. Laffont, F. Lalande, C. Le Roux and J. L. Ogier, An evaluation of the extrinsic cells number in a memory array using cross-correlation products and deconvolution: an instance of a microelectronics experimental inverse problem *Inverse problems in science and engineering (*2011) In press.

[27] H. Bichsel, T. Hiraoka and K. Omata, Aspects of fast ion therapy *Rad. Res.* **153** (2000) 208 – 219.

[28] I. Kawrakow and D.W.O. Rogers, The EGSnrc code system: Monte Carlo simulation of electron and photon transport. *NRCC Report* PIRs-701 NRC, Canada, 2000.